\documentclass{revtex4}

\usepackage{graphicx}
\usepackage{bm}
\usepackage{amssymb}
\usepackage{amsmath}

\begin{document}

\title{Many-body forces with the envelope theory}

\author{Claude \surname{Semay}}
\email[E-mail: ]{claude.semay@umons.ac.be}
\author{Guillaume \surname{Sicorello}}
\email[E-mail: ]{guillaume.sicorello@student.umons.ac.be}
\affiliation{Service de Physique Nucl\'{e}aire et Subnucl\'{e}aire,
Universit\'{e} de Mons,
UMONS Research Institute for Complex Systems,
Place du Parc 20, 7000 Mons, Belgium}
\date{\today}

\begin{abstract}
Many-body forces are sometimes a relevant ingredient in various fields, such as atomic, nuclear or hadronic physics. Their precise structure is generally difficult to uncover. So, phenomenological effective forces are often used in practice. Nevertheless, they are always very heavy to treat numerically. The envelope theory, also known as the auxiliary field method, is a very efficient technique to obtain approximate, but reliable, solutions of many-body systems interacting via one- or two-body forces. It is adapted here to allow the treatment of a special form of many-body forces. In the most favourable cases, the approximate eigenvalues are analytical lower or upper bounds. Otherwise, numerical approximation can always be computed. Two examples of many-body forces are presented, and the critical coupling constants for generic attractive many-body potentials are computed. Finally, a semiclassical interpretation is given for the generic formula of the eigenvalues.
\end{abstract}

\maketitle

\section{Introduction}
\label{sec:intro}

Generally, two-body forces are the only type of interaction considered in many-body quantum systems. But three-body forces (and more generally many-body  forces) are sometimes a crucial ingredient in atomic physics \cite{gatt11}, nuclear physics \cite{ishi17}, or hadronic physics \cite{ferr95,dmit01,pepi02,desp92}. Many-body forces have deep theoretical foundations, but their structure can be very difficult to compute. Effective forms can then be used to take into account at best possible these complicated many-body contributions. Among the possible structures for a $K$-body force in a $N$-body system, one often chosen for its practical use is given by \cite{gatt11,dmit01,pepi02}
\begin{equation}
\label{rK}
\sum_{\{i_1,\ldots,i_K\}}^N V \left( r_{\{i_1,\ldots,i_K\}} \right) \quad \textrm{with} \quad r_{\{i_1,\ldots,i_K\}}^2=\sum_{i < j}^{\{i_1,\ldots,i_K\}} r_{ij}^2 ,
\end{equation}
where $r_{ij}^2 =( \bm r_i-\bm r_j)^2$ and $\{i_1,\ldots,i_K\}$ is a set of $K$ particles among the $N$ possible ones, with $i_1 <\ldots< i_K$. The sum $\sum_{\{i_1,\ldots,i_K\}}^N$ runs over the $C_N^K$ different sets $\{i_1,\ldots,i_K\}$, while the sum $\sum_{i < j}^{\{i_1,\ldots,i_K\}}$ runs over the $C_K^2$ different pairs in a particular set $\{i_1,\ldots,i_K\}$, where $C_A^B$ is a usual binomial coefficient. If $K=2$, the usual two-body case is recovered. Even with a simple phenomenological structure, many-body effects in many-body systems are always very heavy to treat numerically. That is why it is interesting to develop efficient methods to obtain reliable results, even at the price of approximations. 

The envelope theory (ET) \cite{hall80,hall83,hall04}, independently rediscovered under the name of auxiliary field method \cite{silv10}, is a simple technique to compute approximate solutions, eigenvalues and eigenvectors, of many-body systems with arbitrary kinematics in $D$ dimensions \cite{sema13a,sema17}. The basic idea is to replace the Hamiltonian $H$ under study by an auxiliary Hamiltonian $\tilde H$ which is solvable, the eigenvalues of $\tilde H$ being optimised to be as close as possible to those of $H$. Quite good approximations can be obtained for various systems containing up to 10 bosons \cite{sema15a}. The accuracy can be improved, but to the detriment of the possible variational character \cite{sema15b}. The ET can yield interesting results for systems of $N$ identical particles, whose Hamiltonians are given by \cite{sema13a,silv11}
\begin{equation}
\label{HN12}
H=\sum_{i=1}^N T(p_i) + \sum_{i=1}^N U\left(s_i\right) + \sum_{i < j}^N V\left(r_{ij}\right),
\end{equation}
with $p_i = |\bm p_i|$ and $s_i = |\bm r_i - \bm R|$, where $\bm R = \frac{1}{N}\sum_{i=1}^N \bm r_i$ is the centre of mass position. $T$ is the kinetic energy, and $U$ and $V$ are potentials ($\hbar=c=1$). As only the internal motion is relevant, $\sum_{i=1}^N \bm p_i = \bm 0$. The momentum $\bm p_i$ and position $\bm r_i$ of the particle $i$ are conjugate variables. 

The purpose of this work is to generalise ET to treat Hamiltonians with $K$-body forces of type
\begin{equation}
\label{HN1K}
H=\sum_{i=1}^N T(p_i) + \sum_{i=1}^N U\left(s_i\right) + \sum_{\{i_1,...,i_K\}}^N V \left( r_{\{i_1,\ldots,i_K\}} \right).
\end{equation}
Let us note that the Hamiltonian can contain several many-body potentials with various values of $K$. We keep here only one many-body contribution to lighten the demonstration. The one-body term $U$ is kept, because its treatment is a little bit different. 

In Sec.~\ref{sec:ete}, the exact solution for the non-relativistic system of $N$ identical harmonic oscillators with $K$-body forces is first given, and the ET treatment, based on this solution, is then developed for general Hamiltonians. Two analytical examples are presented in Sec.~\ref{sec:examp}. Critical coupling constants for generic attractive many-body potentials are computed in Sec.~\ref{sec:ccc}. A semiclassical interpretation is given for the generic formula of the eigenvalues in Sec.~\ref{sec:semic}. Concluding remarks are given in the last section.
 
\section{Envelope theory equations}
\label{sec:ete}

Let us first consider the following harmonic oscillator type Hamiltonian
\begin{equation}
\label{Hho1}
H_{\textrm{ho}}=\frac{1}{2 \mu}\sum_{i=1}^N p_i^2 + \nu \sum_{i=1}^N  s_i^2 + \rho\sum_{\{i_1,...,i_K\}}^N r_{\{i_1,\ldots,i_K\}}^2.
\end{equation}
It can be rewritten
\begin{equation}
\label{Hho2}
H_{\textrm{ho}}=\frac{1}{2 \mu}\sum_{i=1}^N p_i^2 + \nu \sum_{i=1}^N s_i^2 + \rho\, C_{N-2}^{K-2} \sum_{i<j}^N r_{ij}^2.
\end{equation}
The parts proportional to $\rho$ are identical in (\ref{Hho1}) and (\ref{Hho2}) since
\begin{equation}
\label{C}
C_{N}^{K}\,C_{K}^{2} = C_{N-2}^{K-2}\,C_{N}^{2}.
\end{equation}
An eigenvalue $E_{\textrm{ho}}$ of $H_{\textrm{ho}}$ is given by \cite{silv10}
\begin{equation}
\label{Eho}
E_{\textrm{ho}}= Q\,\sqrt{\frac{2}{\mu}\left(\nu+N\,C_{N-2}^{K-2}\,\rho\right)}
\quad \textrm{with} \quad Q= \sum_{i=1}^{N-1}\left( 2\, n_i+l_i+\frac{D}{2} \right).
\end{equation}

In order to find the eigensolutions of Hamiltonian~(\ref{HN1K}), the auxiliary Hamiltonian $\tilde{H}$ is built
\begin{align}
\label{Htilde}
\tilde{H} &= \left.
\sum_{i=1}^N \left[ \frac{p_i^2}{2\mu_i} + T(G(\mu_i)) - \frac{G^2(\mu_i)}{2 \mu_i} \right] \qquad 
\right\} \sum_{i=1}^N \tilde{T}_i(p_i) \nonumber \\
& \quad \left. + \sum_{i=1}^N \left[ \nu_i s_i^2 + U(I(\nu_i)) - \nu_i I^2(\nu_i) \right]  \qquad 
\right\} \sum_{i=1}^N \tilde{U}_i(s_i) \nonumber \\
& \quad \left. + \sum_{\{\}}^N \left[ \rho_{\{\}} r_{\{\}}^2 + V (J (\rho_{\{\}})) -   \rho_{\{\}} J^2 (\rho_{\{\}} ) \right]  \qquad 
\right\} \sum_{\{\}}^N \tilde{V}_{\{\}}(r_{\{\}}) ,
\end{align}
where the symbol $\{\}$ stands for $\{i_1,...,i_K\}$. Quantities $\mu_i$, $\nu_i$ and $\rho_{\{\}}$ are just c-numbers. The auxiliary functions $G$, $I$ and $J$ are such that
\begin{align}
&G(x) = F^{-1}(x), \quad F(x) = \frac{x}{T^{\prime} (x)}, \label{AuxG} \\
&I(x) = K^{-1}(x), \quad K(x) = \frac{U^{\prime} (x)}{2\, x} , \label{AuxI} \\
&J(x) = L^{-1}(x), \quad L(x) = \frac{V^{\prime} (x)}{2\, x}. \label{AuxJ}
\end{align}
They are assumed to be invertible for $x >0$. The principle of the method is to search for the set of parameters $\alpha_0 = \{ \mu_{i,0}; \nu_{i,0}; \rho_{\{\},0} \}$ which extremises the energy $E$ of a particular eigenstate $| \alpha_0 \rangle$ of $\tilde{H}_0$, which is $\tilde{H}$ evaluated in $\alpha_0$ (the quantum numbers are not indicated to lighten the notations)  
\begin{equation}
\label{partialE}
\left. \frac{\partial E}{\partial \mu_i}\right|_{\alpha_0}
=
\left. \frac{\partial E}{\partial \nu_i}\right|_{\alpha_0}
=
\left. \frac{\partial E}{\partial \rho_{\{\}}}\right|_{\alpha_0} = 0 .
\end{equation}
The procedure is detailed in \cite{silv10}, but the main steps are given here. Using the notation $\langle \cdot \rangle_{\alpha_0} = \langle \alpha_0 | \cdot | \alpha_0 \rangle$, the Hellmann-Feynman theorem applied to the parameter $\mu_i$, for instance, implies
\begin{align}
\label{HFG}
0 = \left. \frac{\partial E}{\partial \mu_i}\right|_{\alpha_0}
&=
\left\langle \frac{\partial \tilde{H}_0}{\partial \mu_{i,0}} \right\rangle_{\alpha_0} \\ \nonumber 
&=
\left\langle 
\frac{G^2(\mu_{i,0}) - p_i^2}{2 \mu_{i,0}^2}
      + G^{\prime}(\mu_{i,0}) \left[ T^{\prime}(G(\mu_{i,0}))
          - \frac{G(\mu_{i,0})}{\mu_{i,0}} \right]
 \right\rangle_{\alpha_0} \nonumber \\
&= \left\langle 
\frac{G^2(\mu_{i,0}) - p_i^2}{2 \mu_{i,0}^2}
\right\rangle_{\alpha_0} ,
\end{align}
thanks to (\ref{AuxG}). Similar calculations finally yield $\langle p_i^2 \rangle_{\alpha_0} = G^2(\mu_{i,0})$, $\langle s_i^2 \rangle_{\alpha_0} = I^2(\nu_{i,0})$ and $\langle r_{\{\}}^2 \rangle_{\alpha_0} = J^2(\rho_{\{\},0})$. As all particles are identical, an eigenstate must be completely (anti)symmetrised. This implies that $\mu_{i,0}=\mu_0$, $\nu_{i,0}=\nu_0$ and $\rho_{\{\},0}=\rho_0$ for all particle numbers \cite{silv10}. So $\alpha_0$ stands now simply for $\{ \mu_0,\nu_0,\rho_0\}$. It is then quite natural to define $p_0^2 = G^2(\mu_0)$ and $r_0^2/N^2 = I^2(\nu_0)$ \cite{sema13a}. The value of $J^2(\rho_{\{\},0})$ must now be computed. It is a matter of combinatorial analysis to show that 
\begin{equation}
\label{si2}
\frac{r_0^2}{N^2} = \langle s_i^2 \rangle_{\alpha_0} = \frac{C_N^2}{N^2} \langle r_{ij}^2 \rangle_{\alpha_0}.
\end{equation}
This is true $\forall\, i, j$ because of the symmetry of the wave-function. Due to (\ref{rK}), $\langle r_{\{\}}^2 \rangle_{\alpha_0}=\sum_{i < j}^{\{\}} \langle r_{ij}^2 \rangle_{\alpha_0}$, so
\begin{equation}
\label{si2}
J^2(\rho_{\{\},0}) = \langle r_{\{\}}^2 \rangle_{\alpha_0}= C_K^2 \langle r_{ij}^2 \rangle_{\alpha_0}= \frac{C_K^2}{C_N^2} r_0^2.
\end{equation}
Using these results, an eigenvalue $E$ of $\tilde{H}$ can be written as a function of $p_0$ and $r_0$
\begin{equation}
\label{EHt0}
E = \langle \tilde{H}_0 \rangle_{\alpha_0}
= N T(p_0) + N U\left(\frac{r_0}{N}\right) + C_N^K V\left(\sqrt{\frac{C_K^2}{C_N^2}} r_0\right) .
\end{equation}

The Hamiltonian $\tilde{H}_0$ can also be written $\tilde{H}_0=H_{ho,0}+B(\mu_0,\nu_0,\rho_0)$ with
\begin{equation}
\label{Ht0}
H_{ho,0} = \frac{1}{2 \mu_0}
\sum_{i=1}^N p_i^2 + \nu_0 \sum_{i=1}^N s_i^2 + \rho_0 \sum_{\{\}}^N r_{\{\}}^2,
\end{equation}
and where $B(\mu_0,\nu_0,\rho_0)$ is a c-number which can be deduced from~(\ref{Htilde}). Applied to the Hamiltonian $\tilde{H}_0$, the generalized virial theorem \cite{luch90} gives
\begin{equation}
\label{virHt0}
\frac{1}{\mu_0} \sum_{i=1}^N \langle p_i^2 \rangle_{\alpha_0}
=  2 \nu_0 \sum_{i=1}^N \langle s_i^2 \rangle_{\alpha_0} + 2\rho_0 \sum_{\{\}}^N \langle r_{\{\}}^2 \rangle_{\alpha_0}, 
\end{equation}
that is to say
\begin{equation}
\label{virHt0}
\frac{N}{\mu_0}p_0^2 =
2\, N \nu_0 \frac{r_0^2}{N^2} + 2\, C_N^K \rho_0 \frac{C_K^2}{C_N^2} r_0^2.
\end{equation}
Definitions (\ref{AuxG})-(\ref{AuxJ}) yielding $\mu_0=F(p_0)$, $\nu_0=K(r_0/N)$ and $\rho_0=L(\sqrt{C_K^2/C_N^2}r_0)$, (\ref{virHt0}) finally reduces to
\begin{equation}
\label{virfin}
N p_0 T'\left(p_0 \right)
=
r_0 U' \left(\frac{r_0}{N} \right) + C_N^K \sqrt{\frac{C_K^2}{C_N^2}} r_0 V' \left(\sqrt{\frac{C_K^2}{C_N^2}} r_0 \right).
\end{equation}
Thanks to (\ref{Eho}), a mean value $\langle H_{ho,0} \rangle_{\alpha_0}$  can be written
\begin{equation}
\label{Hho0}
\langle H_{ho,0} \rangle_{\alpha_0} = 
N\frac{p_0^2}{2\mu_0} + \left(\frac{\nu_0}{N} + C_{N-2}^{K-2} \rho_0 \right) r_0^2
= Q\,\sqrt{\frac{2}{\mu_0}\left(\nu_0+N\,C_{N-2}^{K-2}\,\rho_0\right)}.
\end{equation}
With some algebra, (\ref{virHt0}) and (\ref{Hho0}) implies that $p_0 \, r_0 =Q$. Finally, the set of equations giving an approximate energy of Hamiltonian (\ref{HN1K}) is given by  
\begin{align}
\label{EAFM1}
E &= N\, T(p_0) + N\, U\left(\frac{r_0}{N}\right)
+ C_N^K V\left(\sqrt{\frac{C_K^2}{C_N^2}} r_0\right), \\
\label{EAFM2}
p_0 &= \frac{Q}{r_0} , \\
\label{EAFM3}
N\, p_0 \,T'\left(p_0 \right)
&=
r_0\, U' \left(\frac{r_0}{N} \right) + C_N^K \sqrt{\frac{C_K^2}{C_N^2}}\, r_0\, V' \left(\sqrt{\frac{C_K^2}{C_N^2}} r_0 \right) .
\end{align}
The particular state considered is fixed by the value of $Q$. If $K=2$, the equations in \cite{sema13a} are recovered. Let us note that (\ref{EAFM3}) is also obtained by setting $dE/dr_0 = 0$ with the constraint (\ref{EAFM2}), which shows the extremum character of $E$. As expected, the exact solution~(\ref{Eho}) is recovered for $T(x)$, $U(x)$ and $V(x)$ proportional to $x^2$. 

Following (\ref{Htilde}) and (\ref{AuxI}), and the fact that $I(\nu_0)=r_0/N$, $\tilde{U}$ from $\tilde{H}_0$ can be written (the index $i$ is no longer relevant)
\begin{equation}
\label{Utilde}
\tilde{U}_0(x) = U\left( \frac{r_0}{N} \right) + \frac{N}{2\,r_0} U'\left( \frac{r_0}{N} \right) \left( x^2 - \frac{r_0^2}{N^2} \right).
\end{equation}
It is easy to see that $\tilde{U}_0(r_0/N) = U(r_0/N)$ and $\tilde{U}_0'(r_0/N) = U'(r_0/N)$. So functions $\tilde{U}_0$ and $U$ are tangent in $r_0/N$. This is actually the property on which relies the development of the ET \cite{hall80,hall83}. With similar calculations, one can see that functions $\tilde{T}_0$ and $T$ are tangent in $p_0$, and that functions $\tilde{V}_0$ and $V$ are tangent in $\sqrt{C_K^2/C_N^2}r_0$. If, for instance, the system is such that $\tilde{T}_0 \ge T$, $\tilde{U}_0 \ge U$ and $\tilde{V}_0 \ge V$ for all values of their arguments, than the comparison theorem \cite{sema11} implies that $E$ is an upper bound of the exact eigenvalue. A procedure to verify if such a situation happens is to define three functions $b_T$, $b_U$ and $b_V$ such that 
\begin{equation}
\label{hg}
T(x) = b_T(x^2), \  U(x) = b_U(x^2) \  \textrm{and} \  V(x) = b_V(x^2).
\end{equation}
This procedure relies on the fact that ET functions are tangent to genuine functions. It can be shown that, if $b_T''(x)$, $b_U''(x)$ and $b_V''(x)$ are all concave functions, $E$ is an upper bound \cite{hall83}. Conversely, if all these second derivatives are convex functions, $E$ is a lower bound. If the second derivative is vanishing for one or two of these functions, the variational character is solely ruled by the convexity of the other(s). In the other cases, the variational character of the solution cannot be guaranteed. 

\section{Examples}
\label{sec:examp}

In the following, the generic kinetic energy is considered
\begin{equation}
\label{Tfqm}
T(p) = D_\alpha \, p^\alpha,
\end{equation}
with $D_\alpha >0$ and $\alpha > 0$, in order that $T$ be positive and growing with the modulus of the momentum $p$. Such an operator is, for instance, used in the framework of the fractional quantum mechanics \cite{lask02,wei16}. This form encompasses the non-relativistic case ($D_\alpha =1/(2 m)$ and $\alpha = 2$) and the ultra-relativistic case ($D_\alpha =1$ and $\alpha = 1$). In this last case, $T$ is a phenomenological operator since Hamiltonian~(\ref{HN1K}) is not covariant. Moreover, Hamiltonian~(\ref{HN1K}) can be the mass operator only with appropriate potentials. For instance, it is known that massless particles cannot be bound with an attractive potential vanishing at infinity. With $\alpha \le 2$, an upper bound can be obtained for the energy with appropriate potentials. 

\subsection{Power-law potentials}
\label{sec:plaw}

A first interaction which allows analytical bounds is the the power-law potential
\begin{equation}
\label{Vxpow}
V(r) = a \,\textrm{sgn}(b)\, x^b \quad \textrm{with} \quad a > 0.
\end{equation}
With the kinetic energy (\ref{Tfqm}),  (\ref{EAFM2}) and (\ref{EAFM3}) implies that
\begin{equation}
\label{Vxpow2}
r_0=\left[ \frac{\alpha\, N\, D_\alpha\,Q^\alpha}{C_N^K\,a\,|b|} \left(\frac{C_N^2}{C_K^2}\right)^{b/2} \right]^{1/(b+\alpha)}.
\end{equation}
After some simple algebra, the approximate energy is given by
\begin{equation}
\label{Expow}
E=\textrm{sgn}(b)\,(b+\alpha)\left[ \left(\frac{N\,D_\alpha}{|b|}\right)^b \left(\frac{a\,C_N^K}{\alpha}\right)^\alpha \left(\frac{C_K^2}{C_N^2}\right)^{\alpha\,b/2} Q^{\alpha\,b} \right]^{1/(b+\alpha)}.
\end{equation}
The sign of $E$ must be given by the sign of $b$, so the constraint $b > -\alpha$ appears. With $\alpha=b=2$, the exact solution is found. For $K=\alpha=2$ and $b=-1$, this result coincides with the one in \cite{sema15a}, where the numerical accuracy has been tested. For $K=2$ and $D_\alpha=\alpha=1$, it coincides with a calculation in \cite{silv10}. An upper bound is obtained if $\alpha \le 2$ and $b \le 2$.

\subsection{Exponential potentials}
\label{sec:exppot}

Another interaction which allows analytical bounds is the general exponential potential
\begin{equation}
\label{Vexp}
V(r) = -a \,\exp\left(-b\,r^{\gamma}\right) \quad \textrm{with} \quad a, b, \gamma  > 0.
\end{equation}
With the kinetic energy (\ref{Tfqm}),  (\ref{EAFM2}) and (\ref{EAFM3}) implies that
\begin{equation}
\label{Vexp2}
\frac{\alpha\, D_{\alpha}}{a\, b\, \gamma} \frac{N}{C_N^K} \left(\frac{C_N^2}{C_K^2} \right)^{\gamma/2} Q^{\alpha} = r_0^{\alpha + \gamma} e^{-b \left( \sqrt{C_K^2/C_N^2}\, r_0 \right)^{\gamma}}.
\end{equation}
The solution of this equation is given by the multivalued Lambert $W$ function \cite{corl96}. After some algebra, the approximate energy is given by 
\begin{align}
\label{Eexp}
E &= -a \, C_N^K \exp\left( 
   \frac{\alpha + \gamma}{\gamma} W_0(\delta) \right)
\left[ \frac{\alpha + \gamma}{\alpha} W_0(\delta) + 1 \right], \nonumber \\
& \textrm{with} \quad \delta = 
        -\frac{\gamma}{\alpha + \gamma}
        \left( \frac{\alpha\, b^{\alpha/\gamma} D_{\alpha} }{a\, \gamma}
        \frac{N}{C_N^K} \left(\frac{C_K^2}{C_N^2} \right)^{\alpha/2}
        Q^{\alpha}\right)^{\gamma/(\alpha + \gamma)} .
\end{align}
Bound states, that is to say negative energy solutions, can only be obtained with the branch $W_0$. For $K=\alpha=\gamma=2$, this result coincides with the one in \cite{sema15a}, where the numerical accuracy has been tested. The fact that $E < 0$ and that $-1/e \le \delta < 0$ puts constraints on the global quantum number $Q$. For too high values of the quantum numbers $\{n_i,l_i\}$, no bound state exists. An upper bound is obtained if $\alpha \le 2$ and $\gamma \le 2$. 

\section{Critical coupling constants}
\label{sec:ccc}

For some potentials, as the exponential one, only a finite number of bound states exist. Such an interaction can be written under the form
\begin{equation}
\label{gv}
V(r) = - g\, v(r),
\end{equation}
where $g$ is a positive quantity with the dimension of an energy and $v(x)$ a ``globally positive" dimensionless function vanishing at infinity. The critical coupling constant $g_c(\{q\})$, where $\{q\}$ stands for a set of quantum numbers, is such that the potential admits a bound state with the quantum numbers $\{q\}$ if $g > g_c(\{q\})$. A critical coupling constant can be determined with the system~(\ref{EAFM1})-(\ref{EAFM3}) by setting $E=0$ and searching for the conditions on $g$.

For a $N$-body system with the kinetic part (\ref{Tfqm}) and a $K$-body interaction of the form~(\ref{gv}), the critical constant $g_c$, for a quantum state characterised by the global quantum number $Q$, is given by
\begin{align}
\label{gcK}
g_c &= \frac{1}{x_0^\alpha\, v(x_0)} \frac{N}{C_N^K} \left( \frac{C_K^2}{C_N^2}\right)^{\alpha/2} D_\alpha\,Q^\alpha
, \nonumber \\
0 &= x_0\, v^{\prime}(x_0) + \alpha\, v(x_0) .
\end{align}
$g_c$ is an upper (lower) bound of the genuine constant if the ET energy is an upper (lower) bound of the genuine energy. The variable $x_0$ depends only on the form of the function $v(x)$ and on the power $\alpha$. For $K=\alpha=2$, (\ref{gcK}) coincides with the formula given in \cite{sema13a}, which is in agreement with the results obtained in \cite{rich94}. The critical constant for a one-body interaction is also given in \cite{sema13a}.

\section{Semiclassical interpretation}
\label{sec:semic}

Though the ET is a full-quantum calculation, a semiclassical interpretation of the main equations is possible. This is given in \cite{sema13a} for the cases of one-body and two-body interactions. Independently of the value of $D$, equations (\ref{EAFM1})-(\ref{EAFM2}) describe a system of $N$ particles, each with a momentum $p_0$, located at the vertices of a regular simplex in $N-1$ dimensions, whose circumscribed sphere has a radius $r_0/N$. The distance $e$ between two particles is a constant which is the length of the edge of the simplex, with $e=r_0/\sqrt{C_N^2}$ \cite{coxe73}. So, the quantum mechanics with the symmetrisation procedure predict a geometry for the system which is not possible to achieve in a (semi)classical way in our world when $N>3$. 

Is this interpretation still relevant for $K$-body interactions? With the definition (\ref{rK}) of the arguments of the $K$-body potential, it can be expected that $r_{\{\,\}}^2 = C_K^2 \, e^2$ in the simplex. That is to say $r_{\{\,\}} = \sqrt{C_K^2/C_N^2} \, r_0$, which is exactly the argument of $V$ in (\ref{EAFM1}).

The force $\bm F_{i,\{\,\}}$ acting on the particle $i$ from a particular set of particles $\{\,\}$ is given by
\begin{equation}
\label{derivV}
\bm F_{i,\{\,\}} = - \bm \nabla_{\bm r_i} V \left( r_{\{\,\}} \right) 
= -V' \left( r_{\{\,\}} \right) \bm \nabla_{\bm r_i} r_{\{\,\}}
= -V' \left( r_{\{\,\}} \right) \frac{\sum_{j\ne i}\left(\bm r_i-\bm r_j\right)}{r_{\{\,\}}} 
\end{equation}
if $i \in \{\,\}$, and zero otherwise. In the simplex, due to the symmetry of the shape, only the radial force can contribute. The projection of $\bm r_i-\bm r_j$ on this direction $\hat{\bm n}$ (outward) gives \cite{sema13a} 
\begin{equation}
\label{rijrad}
\left(\bm r_i-\bm r_j\right)\cdot \hat{\bm n} = - e \, \cos \alpha \quad \textrm{with} \quad \cos \alpha = \frac{N}{2 \,\sqrt{C_N^2}},
\end{equation}
for all pairs of particles. If the force is not vanishing, given the constant values of $r_{\{\,\}}$ and $e$, 
\begin{equation}
\label{Fin}
\bm F_{i,\{\,\}} \cdot \hat{\bm n} = V' \left(\sqrt{\frac{C_K^2}{C_N^2}} r_0 \right) \sqrt{\frac{C_N^2}{C_K^2}} \frac{1}{r_0} (K-1) \frac{r_0}{\sqrt{C_N^2}}\frac{N}{2 \, \sqrt{C_N^2}}
\end{equation}
in the simplex. It is a matter of combinatorial analysis to show that the number $i$ is present in $C_{N-1}^{K-1}= C_N^K\,K/N$ sets $\{\,\}$. So, the total radial force acting on a particle $i$ is
\begin{equation}
\label{Fintot}
C_{N-1}^{K-1}\,\bm F_{i,\{\,\}} \cdot \hat{\bm n} = C_N^K\, \sqrt{\frac{C_K^2}{C_N^2}}V' \left(\sqrt{\frac{C_K^2}{C_N^2}} r_0 \right).
\end{equation}
Once multiplied by $r_0$ (coming from the kinetic part \cite{sema13a}), this gives exactly the $K$-body contribution in (\ref{EAFM3}). So, the semiclassical interpretation developed for one-body and two-body  forces is also relevant to $K$-body forces.

\section{Concluding remarks}
\label{sec:cr}

The envelope theory is a very simple method to solve eigenvalue quantum equations for $N$ identical particles in $D$ dimensions \cite{sema13a}, with a reasonable accuracy \cite{sema15a}. It is shown here that the treated potentials can include a special type of many-body forces where the radial variable is a sum of squares of relative two-body distance. In the most favourable cases, analytical bounds of the energy can be obtained, like the two examples studied above. The method can also yield information about the critical coupling constant for attractive wells with a finite number of bound states. At last, a semiclassical interpretation of the method can be done in which the system behaves like a set of particles lying at the vertices of a regular simplex. 

A drawback of the envelope theory is the strong degeneracy inherent to this method. For one-body and two-body interactions, it is possible to correct this by combining the method with the dominantly orbital state method \cite{sema15b}. This can lead to improvements of the energies, but the price to pay is the lost of the possible variational character of the eigenvalues. It could be interesting to test if the mixing of these methods is still possible for many-body forces. It could also be interesting to extend the method to systems with two, or more, different types of particles. Applications of this method are potentially numerous in various domains of physics.


\begin{thebibliography}{99} 

\bibitem{gatt11} M. Gattobigio, A. Kievsky, and M. Viviani, Spectra of helium clusters with up to six atoms using soft-core potentials, Phys. Rev. A \textbf{84}, 052503 (2011)
\bibitem{ishi17} S. Ishikawa, Three-Body Potentials in $\alpha$-Particle Model of Light Nuclei, Few-Body Syst. \textbf{58}, 37 (2017)
\bibitem{ferr95} M. Ferraris, M.M. Giannini, M. Pizzo, E. Santopinto, and L. Tiator, A three-body force model for the baryon spectrum, Phys. Lett. B \textbf{364}, 231 (1995) 
\bibitem{dmit01} V. Dmitra\v{s}inovi\'{c}, Cubic Casimir operator of SU$_\textrm{C}$(3) and confinement in the nonrelativistic quark model, Phys. Lett. B \textbf{499}, 135 (2001) 
\bibitem{pepi02} S. Pepin and Fl. Stancu, Three-body confinement force in hadron spectroscopy, Phys. Rev. D \textbf{65}, 054032 (2002) 
\bibitem{desp92} B. Desplanques, C. Gignoux, B. Silvestre-Brac, P. Gonz\'{a}lez, J. Navarro, and S. Noguera, The baryonic spectrum in a constituent quark model including a three-body force, Z. Phys. A \textbf{343}, 331 (1992) 
\bibitem{hall80} R.L. Hall, Energy trajectories for the $N$-boson problem by the method of potential envelopes, Phys. Rev. D \textbf{22}, 2062 (1980)
\bibitem{hall83} R.L. Hall, A geometrical theory of energy trajectories in quantum mechanics, J. Math. Phys. \textbf{24}, 324 (1983) 
\bibitem{hall04} R.L. Hall, W. Lucha, and F.F. Sch\"oberl, Relativistic $N$-boson systems bound by pair potentials $V(r_{ij}) = g(r^2_{ij})$, J. Math. Phys. \textbf{45}, 3086 (2004)
\bibitem{silv10} B. Silvestre-Brac, C. Semay, F. Buisseret, and F. Brau, The quantum ${\cal N}$-body problem and the auxiliary field method, J. Math. Phys. \textbf{51}, 032104 (2010)
\bibitem{sema13a} C. Semay and C. Roland, Approximate solutions for $N$-body Hamiltonians with identical particles in $D$ dimensions, Res. in Phys. \textbf{3}, 231 (2013)
\bibitem{sema17} C. Semay and F. Buisseret, Bound Cyclic Systems with the Envelope Theory, Few-Body Syst. \textbf{58}, 151 (2017)
\bibitem{sema15a} C. Semay, Numerical Tests of the Envelope Theory for Few-Boson Systems, Few-Body Syst. \textbf{56}, 149 (2015)
\bibitem{sema15b} C. Semay, Improvement of the envelope theory with the dominantly orbital state method, Eur. Phys. J. Plus \textbf{130}, 156 (2015)
\bibitem{silv11} B. Silvestre-Brac and C. Semay, Duality relations in the auxiliary field method, J. Math. Phys. \textbf{52}, 052107 (2011)
\bibitem{luch90} W. Lucha, Relativistic Virial Theorems, Mod. Phys. Lett. A \textbf{5}, 2473 (1990)
\bibitem{sema11} C. Semay, General comparison theorem for eigenvalues of a certain class of Hamiltonians, Phys. Rev. A \textbf{83}, 024101 (2011)
\bibitem{lask02} N. Laskin, Fractional Schr\"odinger equation, Phys. Rev. E \textbf{66}, 056108 (2002)
\bibitem{wei16} Y. Wei, Comment on ``Fractional quantum mechanics" and ``Fractional Schr\"odinger equation", Phys. Rev. E \textbf{93}, 066103 (2016) 
\bibitem{corl96} R.M. Corless, G.H. Gonnet, D.E.G. Hare, D.J. Jeffrey, and D.E. Knuth, On the Lambert W Function, Adv. Comput. Math. \textbf{5}, 329 (1996) 
\bibitem{rich94} J.-M. Richard and S. Fleck, Limits on the Domain of Coupling Constants for Binding $N$-Body Systems with No Bound Subsystems, Phys. Rev. Lett. \textbf{73}, 1464 (1994)
\bibitem{coxe73} H.S.M. Coxeter. \emph{Regular polytopes} (New York, Dover Publications, 1973)


\end{thebibliography}
\end{document}